\documentclass[prc,aps,amsfonts,twocolumn,dvips]{revtex4}
\usepackage{graphics,epsfig,amssymb}

\begin{document}

\title{\bf
Generic strong coupling behavior of Cooper pairs in the surface of superfluid
nuclei
}

\author{\rm N. Pillet$^{(a)}$, 
N. Sandulescu$^{(a,b,c)}$, 
P. Schuck$^{(c,d)}$ }

\bigskip

\address{\rm
$^{(a)}$DPTA/Service de Physique nucl\'eaire, CEA/DAM Ile de France, 
BP12, F-91680 Bruy\`eres-le-Ch\^atel, France \\
$^{(b)}$ Institute of Physics and Nuclear Engineering, 76900 Bucharest,
Romania \\
$^{(c)}$~  Institut de Physique Nucl\'eaire, CNRS, UMR8608, Orsay,
F-91406, France \\
$^{(d)}$~  Universit\'e Paris-Sud, Orsay,
F-91505, France }
\date{\today}

\def\fid{\vert\phi >}
\def\fig{< \phi\vert}
\def\psid{\vert\Psi>}
\def\psig{<\Psi\vert}
\def\psid{\vert\Psi>}
\def\psig{<\Psi\vert}
\def\dspt{\displaystyle}

\begin{abstract}
With realistic HFB calculations, using the D1S Gogny force, we 
reveal a generic behavior of concentration of small sized Cooper pairs 
(2-3 fm) in the surface of superfluid nuclei. This study confirms and extends 
previous results given in the literature that use more schematic 
approaches. 
\end{abstract}

\maketitle

The opportunities offered by the new radioactive beam facilities to study the 
properties of weakly bound nuclei with large neutron skins or halos triggered 
new interest for the issue of space correlations induced by the formation of 
Cooper pairs. The spatial correlations 
of Cooper pairs in superfluid nuclei have not been extensively studied in the past, but 
nevertheless a certain number of investigations, some rather early, do exist. 
Mostly this was done for the single Cooper pair problem. For example the 
rms diameter of the extra neutron pair in $^{18}$O is shown as a function of 
the nuclear radius by Ibarra et al.~\cite{Ibarra}. One sees a strong 
minimum in the nuclear surface, indicating an rms separation between the two 
active neutrons of the order of 2-3 fm. A similar behaviour was found later
by Catara et al.~\cite{Catara} and Ferreira et al.~\cite{Ferreira} for
a neutron pair in $^{206}$Pb and $^{210}$Pb. More recently, there are also many investigations of the single
Cooper pair problem in the halo nucleus $^{11}Li$~\cite{Bertsch,Hagino}. \\
One of the rare papers 
where spatial correlations of Cooper pairs are investigated in superfluid 
nuclei is the one of Tischler et al.~\cite{Tischler} where  
the probability distribution of the pairs is shown as a function of the 
center of mass 
$R=\frac{1}{2}|{\vec r}_1 + {\vec r}_2|$ and the relative distance of the 
nucleons in the
pairs $r= |{\vec r}_1-{\vec r}_2|$ with (${\vec r}_1,{\vec r}_2$) the coordinates of two
nucleons. They showed that in the open shell isotope $^{114}$Sn
one also finds Cooper pairs with short range space correlations, like in 
one pair systems. They confirmed also the finding of Catara et al, i.e., the important 
role played by the parity mixing for inducing short range space correlations. 
Most of those older works were, however, done using rather schematic 
models and/or pairing forces. There exists, however, one study 
with a realistic pairing force (i.e., the Gogny interaction) by 
Barranco et al.~\cite{Barranco}, dedicated to nuclei embedded in a 
neutron gas, a system found in the inner crust of neutron stars.
One of the first systematic analyses of strong di-neutron spatial correlations  induced 
by the pairing interaction was done recently by Matsuo et al.\cite{Matsuo},
using a zero range pairing force. 
The study of nuclear surface pairing properties was also the aim of several 
half infinite matter investigations~\cite{Farine}, ~\cite{Baldo}. It was 
found that the pair density reaches out further than the ordinary density but 
neither the local coherence length nor the probability distribution of the 
pairs were calculated.

The aim of the present work is to verify how much all these relative 
scattered pieces of information withstand a general study of superfluid nuclei 
using one of the most performant HFB approaches, that is employing the finite 
range Gogny D1S-interaction~\cite{D1S}. As a matter of fact we will see that many of the 
earlier findings are qualitatively or even quantitatively confirmed. Indeed, we  
will show that the strong concentration of pair probability of small Cooper pairs in the nuclear 
surface is a quite general and generic feature and that nuclear pairing 
is much closer to  the strong coupling regime~\cite{Matsuo,Baldo1} than 
previously assumed.\\
We will start by explaining shortly how the spatial properties of nuclear pairing 
are investigated within the HFB approach and then we shall present our results and 
conclusions. For further understanding of the phenomena, a simple 
semiclassical analytic model for nuclear pairing will be also considered. \\

It is well known \cite{ring} that pairing correlations can be adequately studied
with the Cooper-pair probability $|\kappa|^2$, where (in standard notation): 
\begin{equation}
{\kappa}_{\sigma,\sigma'}({\vec r}_1,{\vec r}_2)= 
<HFB|{\psi}_{\sigma'}({\vec r}_2){\psi}_{\sigma}({\vec r}_1)|HFB>~,
 \end{equation}
 is the anomalous density matrix or
pairing tensor in $r$-space, calculated with the HFB-wave function.
Following Refs.~\cite{Catara,Tischler}, we shall also consider
the quantity: 
\begin{equation}
P(R,r)= R^2r^2|\kappa(R,r)|^2,
\label{eq2} 
\end{equation}
which is the pair probability averaged over the angle between ${\vec R}$ and $\vec r$ and 
multiplied by the phase space factors $R^2r^2$. This quantity is important since it determines the
two-particle spectroscopic factor \cite{catara2} and other expectation values of two-body operators
(e.g., pairing energy).
 Let us mention that Eq.(\ref{eq2}) is 
formally the same for the single Cooper pair problem~\cite{Ibarra,Catara,Ferreira,Hagino} 
and for the case of Cooper pairs in a condensate~\cite{Tischler}. \\
In what follows, we shall consider the HFB expression of the pairing tensor  $\kappa$ 
in center of mass and relative coordinates given by:
\begin{equation}
\begin{array}{c}
\dspt \kappa(\vec R,\vec r)=\frac{1}{4\pi}\sum_{n_1,n_2,l_1j_1} \kappa_{n_2,n_1}^{l_1j_1} ~~~~~~~~~~~~~~~~\\
\dspt \times \sum_{nNl} (-)^l (\frac{2l+1}{2l_1})^{1/2} u_{nl}(r/\sqrt2) u_{Nl}(\sqrt2R) \\
\dspt \times P_l(cos\hat{rR}) <nlNl;0|n_1l_1n_2l_1;0>~,
\end{array}
\end{equation}
where $<nlNl;0|n_1l_1n_2l_1;0>$ is the Brody-Moshinski bracket, 
$u_{nl}(r)$ are the radial wave functions of the harmonic oscillator 
and $\kappa_{n'n}^{lj}$ is the matrix of the pairing tensor for a given angular
momentum $lj$. As defined here, the latter has an intrinsic parity $(-)^l$.
\begin{figure}
\vspace{-1.2cm}
\includegraphics[height=13.0cm,angle=0]{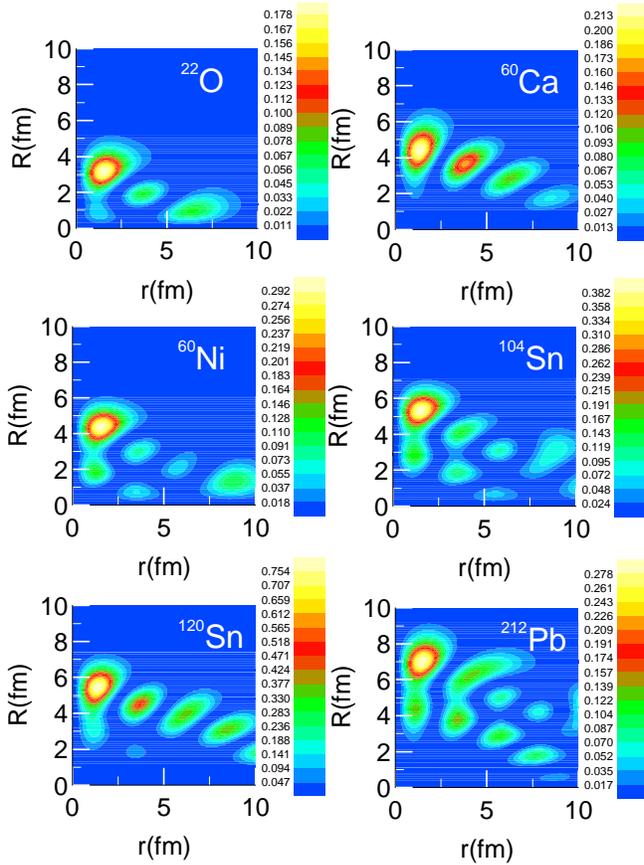}
\vspace{-1.5cm}
\caption{$P(R,r)$ calculated with HFB-D1S for $^{22}O$,$^{60}Ca$, $^{60}Ni$, 
$^{104}Sn$, $^{120}Sn$, $^{212}Pb$. Scales have been multiplied by a factor of $10^2$.}
\label{figure2}
\end{figure}
The HFB calculations are performed with the D1S Gogny
force ~\cite{Decharge}. In the calculations a basis with 15 
harmonic oscillator shells have been considered. 
The contour-lines of $P(R,r)$ for various  
superfluid nuclei are shown in FIG.\ref{figure2}. The striking feature is 
that for all these nuclei the same scenario, with only slight modulations, 
emerges: the probability $P(R,r)$ is strongly concentrated in the surface with a small
diameter of the pairs of the order of $2-3 fm$. 
In FIG.\ref{figure2}, we show nuclei close to the neutron drip-line ($^{60}$Ca) as well as nuclei
closer to stability. Seemingly, there is no essential difference in the behavior of
$P(R,r)$ between very neutron-rich nuclei and stable ones. This fact explains why one finds in
all the superfluid nuclei a high probability for two-neutron transfer reactions. \\
\begin{figure}
\vspace{-2.0cm}
\includegraphics[height=10.0cm,angle=0]{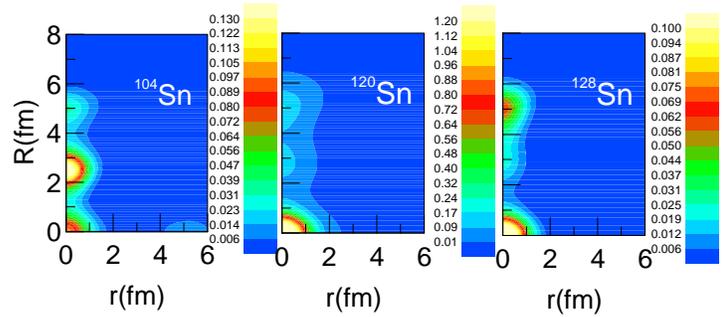}
\vspace{-4.5cm}
\caption{$|\kappa(R,r)|^2$ calculated with HFB-D1S for $^{104}Sn$, $^{120}Sn$, $^{128}Sn$.
Scale has been multiplied by a factor of $10^{6}$.}
\label{figure7}
\end{figure}
To conclude from the strong concentration of $P(R,r)$ in the surface that Cooper pairs are mostly sitting
in the surface of the nucleus may be, however, a bit misleading. In FIG.\ref{figure7}, we show the pair probability
$|\kappa(R,r)|^2)$ without the factor $R^2r^2$ which enters in the probability $P(R,r)$.
One can notice that for all tin isotopes
the Cooper pairs have very small extension in r-direction throughout the nuclear radius. 
 However, 
the distribution in $R$ is rather different in the three isotopes.
The difference comes from the localisation properties of the single-particle
shells which are closest to the chemical potential. Thus, the pronounced 
concentration of $|\kappa(R,r)|^2$ around R=5 fm in $^{128}Sn$ is due to the surface localisation
of the single-particle wave function $1h_{11/2}$, which becomes much closer
to the chemical potential in this isotope compared to lighter ones. One can also notice
that in $^{120,128}Sn$ the pair probability has a sizeable  value for small
values of $R$, which comes mainly from the contribution of the state 
$3s_{1/2}$ to pairing correlations. If we have had chosen the neutron deficient Pb-isotopes, in
which there is no $s$-state in the major shell, one would rather see a depression 
of pair probability at the origin. Therefore, to say where the Cooper pairs are 
preferentially located in nuclei is a somewhat subtle question because the answer depends 
rather strongly on the shell structure (see also \cite{ssv}). 
The shell structure dependence of  $|\kappa(R,r)|^2$ is 
largely washed out by the phase factor $r^2R^2$ when $P(R,r)$ is calculated. 
We shall make further investigations on this issue in a future work.
However, we want to point out again that in all expectations values, like e.g. the
pairing energy and spectroscopic factors, it is $P(R,r)$ which counts and not $|\kappa(R,r)|^2$.  

\begin{figure}
\vspace{-1.1cm}
\hspace{-1.0cm}
\includegraphics[height=7.0cm,angle=0]{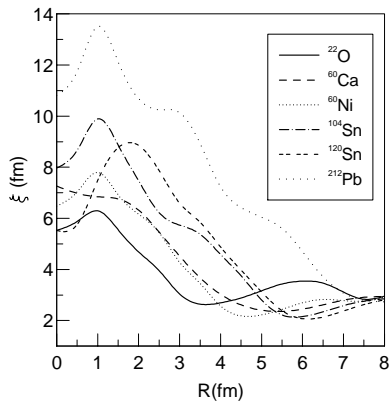}
\vspace{-0.8cm}
\caption{ Coherence length $\xi(R)$ for $^{22}O$,$^{60}Ca$, $^{60}Ni$, 
$^{104}Sn$, $^{120}Sn$, $^{212}Pb$.}
\label{figure6}
\end{figure}

The fact that the Cooper pairs with small size are concentrated in the surface 
can be also seen from the dependence of the coherence length on the center of 
mass of the pairs. The coherence length is defined as: 
\begin{equation}
\dspt \xi(R)= \frac{(\int r^4 |\kappa(R,r)|^2 dr)^{1/2}}
                 {(\int r^2 |\kappa(R,r)|^2 dr)^{1/2}}
\label{se1}
\end{equation} 
It is shown for various nuclei in FIG.\ref{figure6}. 
One sees well defined and pronounced minima at $\xi \sim ~2-3 fm$ for R of the order
of the surface radius. 
 As we have already mentioned, a small coherence length in the case of a single Cooper 
pair has 
already been found previously for $^{18}O$ in Ref.\cite{Ibarra}. It is also
the case for the Cooper pair in $^{11}Li$~\cite{Bertsch,Hagino}.
Our calculations did not allow to go 
much beyond the minima because of the employed  harmonic oscillator basis what 
becomes inaccurate far outside the nuclear radius. 
However, the position of the minima 
is always clearly identified and seen to be similar in all cases.
What is surprising is that the size of the Cooper pairs 
starts to decrease already well inside, around $R=2 fm$. 
Moreover, the decrease towards the surface is approximately linear.\\

\begin{figure}
\vspace{-0.9cm}
\hspace{-2.2cm}
\includegraphics[height=7.0cm,angle=0]{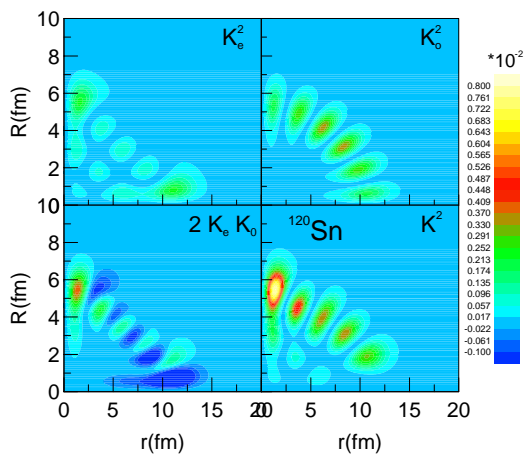}
\vspace{-0.5cm}
\caption{Contributions of definite parity to $P(R,r)$ calculated with HFB-D1S for $^{120}Sn$.}
\label{figure3}
\end{figure}

In order to demonstrate that the strong concentration of small Cooper pairs
in the surface of the nuclei is not a trivial effect, we decompose $\kappa(R,r)$
in a part $\kappa_e(R,r)$ which contains only even parity wave functions and a part $\kappa_o(R,r)$
which contains only the odd parity ones, i.e., $\kappa(R,r) = \kappa_e(R,r) + \kappa_o(R,r)$.
In FIG.\ref{figure3}, we show what are the probability distributions for $P_e(R,r)$, $P_o(R,r)$ 
and $P_{eo}(R,r)$ in the case of $^{120}Sn$. The quantity $P_{eo}(R,r)$ corresponds to 
the interference term $2 \kappa_e \kappa_o$. From FIG.4 one can see that selecting
only either even or odd parity states in $\kappa(R,r)$ has a strong 
delocalisation effect on the Cooper pairs: they are democratically distributed with respect to an
interchange of R and r variables (one should notice that in Eq.(3) the symmetry between $R$
and $r$ involves a factor 2, which comes through the standard definition of Brody-Moshinsky
transformation). So no small Cooper pairs in the  nuclear surface are prefered at 
all in those cases. The concentration only shows up, when even and odd parity states are mixed. 
This is clearly revealed in looking at the interference term $P_{eo}(R,r)$. We see that it is 
negative for regions close to the r-axis and positive close to the R-axis. We checked that this 
scenario stays the same for all other superfluid nuclei considered. This scenario was also
nicely described in the papers by Catara et al.~\cite{Catara} and Tischler et al.~\cite{Tischler}. Mixing of parities naturally occurs in heavy nuclei because of the presence of intruder states of unatural parity in the main shells of given parity. However, as seen in FIG.\ref{figure2} first panel, the concentration of Cooper pairs also occurs in such light nuclei as the oxygen isotopes where no intruders are present. This means that pairing in nuclei is sufficiently strong so that $\kappa$ grabs contributions from several main shells, allowing for parity mixing even in light nuclei. If one artificially restricted the pairing configurations, e.g. in $^{22}O$, to 
the s-d shell, then certainly no concentration effect at all would be seen 
(in this respect, see also the study in ~\cite{Tischler}). 
So to grasp the full physics of nuclear pairing it is very 
important to work in a large configuration space, 
comprising several shells below and above the active one (see also 
~\cite{Matsuo}). \\
In order to understand in more detail where this extraordinary 
concentration effect from even-odd parity mixing comes from, let us consider a very 
simple model. We got inspired by the Thomas-Fermi model 
presented in Ref.\cite{Ferreira} where the anomalous density matrix is given by
${\kappa}^{TF}(R,r) \sim k_F(R) j_0(k_F(R)r)$
with $j_0(x)$ a spherical Bessel fonction, 
$k_F(R)=\sqrt{\frac{2m}{\hbar^2}(\mu-V(R))}~ \Theta(\mu-V)$ the local Fermi momentum, 
$\mu$ the chemical potential (or Fermi energy), and $V(R)$ is 
phenomenological mean field potential.
It can be shown~\cite{Vinas} that a slightly more 
elaborate semiclassical version can be written as
\begin{equation}
{\kappa}^{sc}(R,r)= \frac{m}{\hbar^2 \pi^2} \int~dE~\kappa(E)~k_E(R)~j_0(k_E(R)r)~,
\label{eq5} \end{equation}
where $k_E(R)$ is the local momentum at energy E, obtainable from $k_F$ in 
replacing $\mu$ by $E$, and $\kappa(E)$ is the continuum version of the 
$\kappa$'s for the individual quantum levels:
$\kappa(E)= \Delta(E)/(2 \sqrt{(E-\mu)^2+\Delta(E)^2})$. 
We see that for very small $\Delta's$, one 
gets back the TF model \cite{Ferreira}. However, for realistic gap values the 
distribution of $\kappa$'s is very important, otherwise the concentration effect
will not show up.
For $\Delta(E)$ we adjust a Fermi function to represent on average the gap-
values of the individual single particle levels. An example can be seen on 
FIG.\ref{figure3} of Ref.\cite{Taruishi}. In the present work, we have fitted the function 
$\Delta(E)$ on HFB-D1S results for $^{120}Sn$.
A good fit function is given by 
$\Delta(E)= 4/[1+exp(E-\mu)/20]$ (all numbers are in MeV). For the mean field potential $V(R)$ we 
take the Woods-Saxon form of Ref.~\cite{Shlomo}. The chemical potential $\mu$ is 
determined, as usual, via the particle number condition. \\
\begin{figure}
\vspace{-1.2cm}
\hspace{-2.2cm}
\includegraphics[height=8.0cm,angle=0]{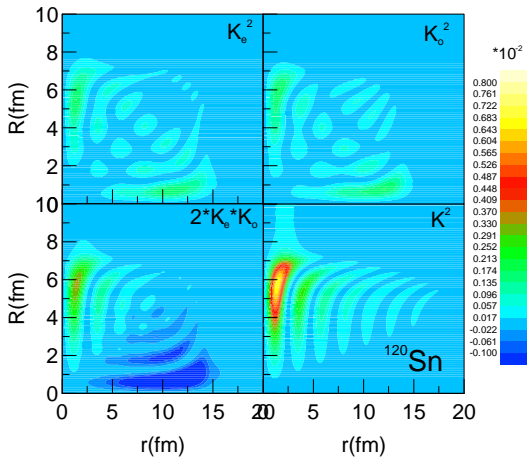}
\vspace{-1.0cm}
\caption{Parities contributions $P^{sc}(R,r)$ calculated with the semiclassical model for $^{120}Sn$.}
\label{figure1}
\end{figure}
In FIG.\ref{figure1} we show the 
corresponding semiclassical probability ${P}^{sc}(R,r)$. We see qualitatively  good 
agreement with the quantal HFB-results, for instance in what concerns the concentration of
small Cooper pairs in the surface. We also show in FIG.\ref{figure1} the parity projected probabilities.
As in FIG.\ref{figure3}, one sees the strong delocalisation effect. In our model this can be understood 
analytically. Parity projection can be written as
$\kappa_{e/o}({\vec r}_1,{\vec r}_2)= 
\frac{1}{2}[\kappa({\vec r}_1,{\vec r}_2) \pm \kappa({\vec r}_1,-{\vec r}_2)]=
\frac{1}{2}[\kappa({\vec R},{\vec r}) \pm \kappa({\frac{\vec r}{2}},2{\vec R})]$ (see Ref.~\cite{Vinas1}).
We therefore see that good parity implies, up to a scale factor, 
a symmetrisation in coordinates $R$ and $r$. 
This is general and can be investigated analytically in the TF model.
We also calculate the coherence length semiclassicaly. 
We find qualitatively the same behavior as in the quantal calculation of FIG.\ref{figure6}. \\
The analytic model also allows to quickly grasp the significance 
of the coordinates used by Matsuo et al.~\cite{Matsuo}. 
There, one  takes a reference particle at position $\vec r_1$ on the z-axis, 
i.e. $\vec r_1=z_1 \vec e_z$. Moving the second particle on the z-axis 
we see that for $P_{e/o}$ two symmetric peaks 
at $\vec r_2=z_1 \vec e_z$ and $\vec r_2=-z_1 \vec e_z$ appear 
whereas for the total probability only one peak on the side of 
the test particle appears. This is a clear signature of strong pairing 
correlations as also pointed out in ~\cite{Matsuo}. 

In conclusion, we showed that Cooper pairs in superfluid nuclei preferentially 
are located with small size ($2-3 fm$) in the surface region. There, they 
maximally profit from the Cooper phenomenon, that is, with respect to the 
neutron-neutron virtual S-state in the vacuum (rms 12 fm, \cite{Hagino}), strong 
extra binding occurs, 
as long as the density is not too high. Further to the center of the nucleus 
the stronger effect of the orthogonalisation of the pair with respect to the 
denser core-neutrons perturbs the pair wave function. It starts to oscillate 
and expands again~ \cite{Hagino}. That this simple, physically appealing and generic picture, 
is so pronounced, has come as a surprise. It is certainly important for the interpretation of pair transfer reactions. Most of these facts had already been 
revealed in the past for specific examples and schematic models and forces. 
We think, it is the merit 
of this work that it demonstrates with realistic HFB calculations using the finite range D1S force the generic aspect of strong coupling features of singlet isovector pairing in nuclei. 
These features are in agreement with the ones recently put forward by Matsuo et al.~ \cite{Matsuo}. 
Let us mention that the strong coupling features 
revealed here are somewhat contrary to the old believe~\cite{Bohr-Mott} that the  
coherence length of nuclear pairs is 
of the same order or larger than the nuclear diameter. On the contrary, a much 
more diverse local picture has emerged. This may also be the reason for the 
rather good succes of LDA for nuclear pairing found in the past~\cite{Kucharek}.
In spite of the strong coupling aspects revealed in this work, we hesitate to 
say that there is Bose-Einstein condensation (BEC) of isovector Cooper pairs, 
since this, strictly speaking, occurs only for ( in infinite matter) negative 
chemical potential, what means true binding. However, $\mu$ never turns 
negative for isovector pairing in infinite nuclear or neutron  matters.
Nuclear isovector pairing is just in the transition region from BEC to BCS. \\
Acknowledgements : We would like to thank Marc Dupuis for his help in the HFB code.

\end{document}